# AN ENDURING RAPIDLY MOVING STORM AS A GUIDE TO SATURN'S EQUATORIAL JET'S COMPLEX STRUCTURE

A. Sánchez-Lavega[1], E. García-Melendo[1-2], S. Perez-Hoyos[1], R. Hueso[1], M. H. Wong[3], A. Simon[4], J. F. Sanz-Requena[5], A. Antuñano[1], N. Barrado-Izagirre[1], I. Garate-Lopez[1], J. F. Rojas[1], T. del Rio Gaztelurrutia[1], J. M. Gómez-Forrellad[2], I. de Pater[3], L. Li[6], T. Barry[7] and PVOL contributors[8]

[1] Universidad del País Vasco UPV/EHU, Escuela de Ingeniería de Bilbao, Departamento Física Aplicada I, Alameda Urquijo s/n, 48013 Bilbao. Spain
[2] Fundació Observatori Esteve Duran, c/ Montseny, 46 – Urb. El Montanyá, 08553 Seva, Barcelona. Spain
[3] University of California, 501 Campbell Hall, Berkeley CA 94720. USA
[4] NASA Goddard Space Flight Center/690, 8800 Greenbelt Rd., Greenbelt, MD 20771. USA
[5] Universidad Europea Miguel de Cervantes, C/Padre Julio Chevalier, 2, 47012 Valladolid. Spain
[6] Department of Physics, University of Houston, Houston, Texas, 77204. USA
[7] Broken Hill Observatory, 406 Bromide St, Broken Hill, NSW 2880. Australia
[8] PVOL contributors listed in Supplementary Information

Correspondence and requests for materials should be addressed to A.S.-L.: (agustin.sanchez@ehu.eus).

**Saturn has an intense and broad eastward equatorial jet with a complex three-dimensional structure mixed with time variability. The equatorial region experiences strong seasonal insolation variations enhanced by ring shadowing and three of the six known giant planetary-scale storms have developed in it. These factors make Saturn's equator a natural laboratory to test models of jets in giant planets. Here we report on a bright equatorial atmospheric feature imaged in 2015 that moved steadily at a high speed of 450 ms$^{-1}$ not measured since 1980-81 with other equatorial clouds moving within an ample range of velocities. Radiative transfer models show that these motions occur at three altitude levels within the upper haze and clouds. We find that the peak of the jet (latitudes 10°N to 10°S) suffers intense vertical shears reaching +2.5 ms$^{-1}$ km$^{-1}$, two orders of magnitude higher than meridional shears, and temporal variability above 1 bar altitude level.**

At the upper cloud level, the giant planets Jupiter and Saturn display a permanent system of alternating eastward and westward zonal jets whose intensity and width show few temporal changes since the first detailed measurements in 1979-1980 [1-9]. One exception is Saturn's broad equatorial jet that extends from planetographic latitudes ~35°N to 35°S reaching eastward peak velocities ~ 450 – 500 ms$^{-1}$ [7-9] (Figure 1). The velocity field that traces the equatorial jet is much more complex than at other latitudes [15], showing a vertical structure and temporal variability that have been so far not well characterized since a key part of the measurements at different altitudes were obtained in different epochs [10-14]. The Equatorial jets of Jupiter and Saturn are particularly relevant to atmospheric dynamics because of their eastward flow (contrary to the equatorial westward flows that occur in other rapidly rotating planets, as Earth, Mars, Uranus and Neptune) and because of their intensity (Saturn Equatorial peak velocities reach ~ 1/3$^{rd}$ of the sound speed). The origin of Jupiter's and Saturn's jets is not well understood, and it is an open issue if they are deep or shallow in vertical extent and if they have deep or shallow forcing sources, or a mixture of both (solar heating, internal energy and latent heat release) [1-2,16]. Distinguishing between these possibilities requires a good characterization of the energy balance in the atmosphere, a quantification of the contribution of each of the above sources to forcing, a precise determination of the wind field, and knowledge of the structure of the deep atmosphere by gravity field measurement [1-2, 16-17].

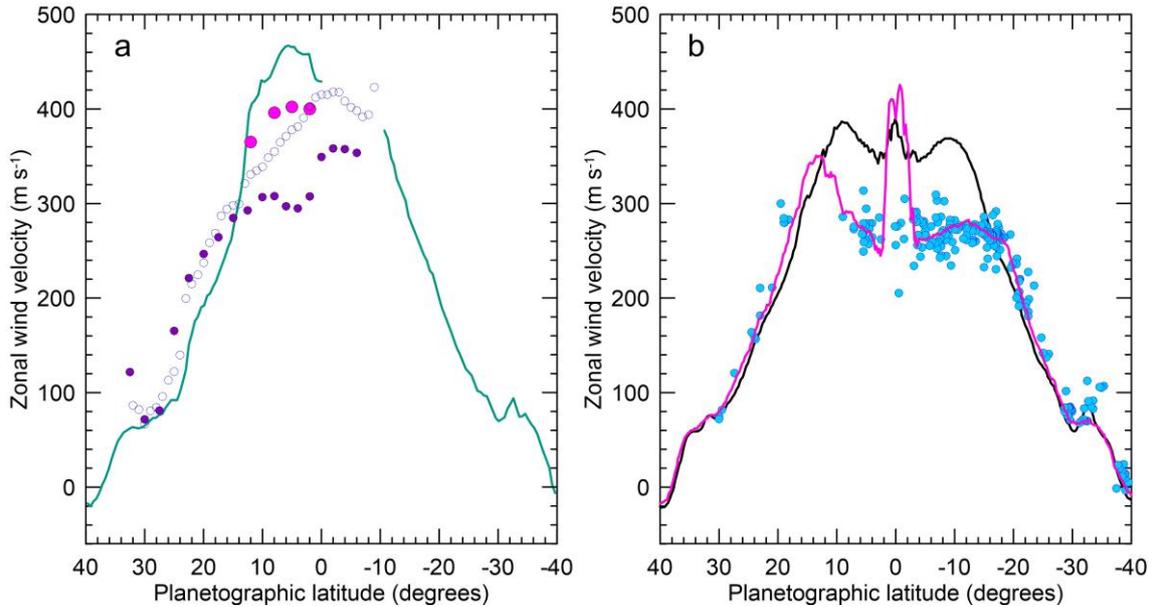

**Figure 1. Saturn's Equatorial jet in time traced by cloud motions.** *(a) Voyager 1 and 2 in 1980-81 (turquoise line, [7]); Hubble Space Telescope (HST) in 1990-91 in the 889 nm methane absorption band (violet filled circles) and in 547 nm (empty circles) corresponding to the development of the 1990 GWS [20]; Ground-based historical GWS storms in 1876, 1933 and 1990 (red dots) [22] (b) Cassini ISS in 2004-09 in the methane band at 889 nm (magenta line, [9, 12]), and in 752 and 939 nm (black line, [9, 12]); HST in 1994-2003 in the 889 nm methane band (blue dots) [10].*

The winds are measured by tracking features in the upper hazes and clouds close to the tropopause (at the altitude pressure level $P \sim 100$ mbar) and upper troposphere ($P \sim 1-4$ bar) where most solar radiation is deposited [18]. This layer has a thickness of ~ 100 - 150 km and acts as the coupling between the deep troposphere and the stratosphere. Interestingly, the equatorial region of Saturn experiences strong seasonal insolation variations enhanced by ring shadowing periods [18]. In addition, Saturn's Equator has been the place of three of the six known Great White Spot (GWS) events that have been observed in the last 134 years [19-22]. These giant planetary-scale storms influence the zonal winds giving insight on jet stability and forcing mechanisms [23-24]. Finally, a semiannual oscillation (SAO) in the temperature and wind fields (occurring in the stratosphere between latitudes 15°N and 15°S) and its role in the upper troposphere represents another open issue of Saturn's equatorial dynamics [25-26]. All these factors make Saturn's equator a natural laboratory where to test models of jet stability and generation in giant planets, a major open issue in geophysical fluid dynamics [1-2, 16], whose implications extend to the case of gas giant exoplanets [27].

Here we address the vertical structure and temporal variability of the puzzling Equatorial jet. Using ground-based and Hubble Space Telescope (HST) images obtained in 2015 in the visual range, we were able to measure wind speeds using cloud tracking at three different altitude levels in the peak of the Equatorial jet between latitudes ~20°N and 20°S. The altitude of the tracers was retrieved using radiative transfer modeling of the spectral and geometrical dependence of the absolute reflectivity across the Equatorial Zone (EZ). By combining these wind measurements and altitude determinations with those previously reported during the last three decades, we constrain the temporal variability of the jet at different altitudes from 1980 to the present, encompassing more than one Saturn's year (29.5 years). Finally we discuss these results in the context of the Equatorial dynamics.

## Results

**Cloud morphology and long-term motion.** Ground-based observations of Saturn obtained during the first half of 2015 showed the presence of a conspicuous bright white spot (WS) at red wavelengths (~ 610 – 950 nm) in the Equatorial Zone (planetographic latitude 6°N). The spot was easily detected by observers using telescopes in the range of 25-40 cm in diameter contributing to the PVOL database of planetary images [28] (Supplementary Table 1, [29]) and with PlanetCam instrument on the 2.2 m telescope at Calar Alto Observatory [30] (Figure 2a). Its motion, tracked from April to October, revealed a steady linear drift in longitude relative to the rotating reference frame System III [31] with a rapid zonal velocity of $u = 447 \pm 1$ ms$^{-1}$ (Figure 2b). The bright feature was accompanied by a dark spot (DS) moving at the same speed and 2º to its East (Figure 3). A similar bright spot was found in PVOL images in July-August 2014, close to the predicted position, moving rapidly with a velocity of $u = 445 \pm 1$ ms$^{-1}$. The drift rate in the longitudinal position of both spots in System III was steady at a rate of -37.05°/day (for the WS in 2015). To determine if these two similar features are in fact the same one tracked on different periods of time we used a reference system in which the 2015 spot is nearly stationary (Figure 2b). The long-term linear drifts of both spots in System III longitude allows to determine the mean speed of both spots with an accuracy of $\pm 1$ ms$^{-1}$. With the data in hand, we cannot definitively conclude if the 2014 WS changed slightly its drift velocity (being the same as the 2015 WS feature), or if these are different spots which emerged separated in time but at close longitudinal positions. In both cases, these speeds are well above the 380 ms$^{-1}$ measured during the Cassini epoch at this latitude [8-9]. These high speeds at the Equator were only observed in 1980-1981 [7] during the Voyager flybys (Figure 1), suggesting one of the following possibilities: the spot is at a deeper level where faster winds are present, [14, 32] or a change in the jet system have occurred since the last accurate wind measurements based on Cassini ISS data in 2009.

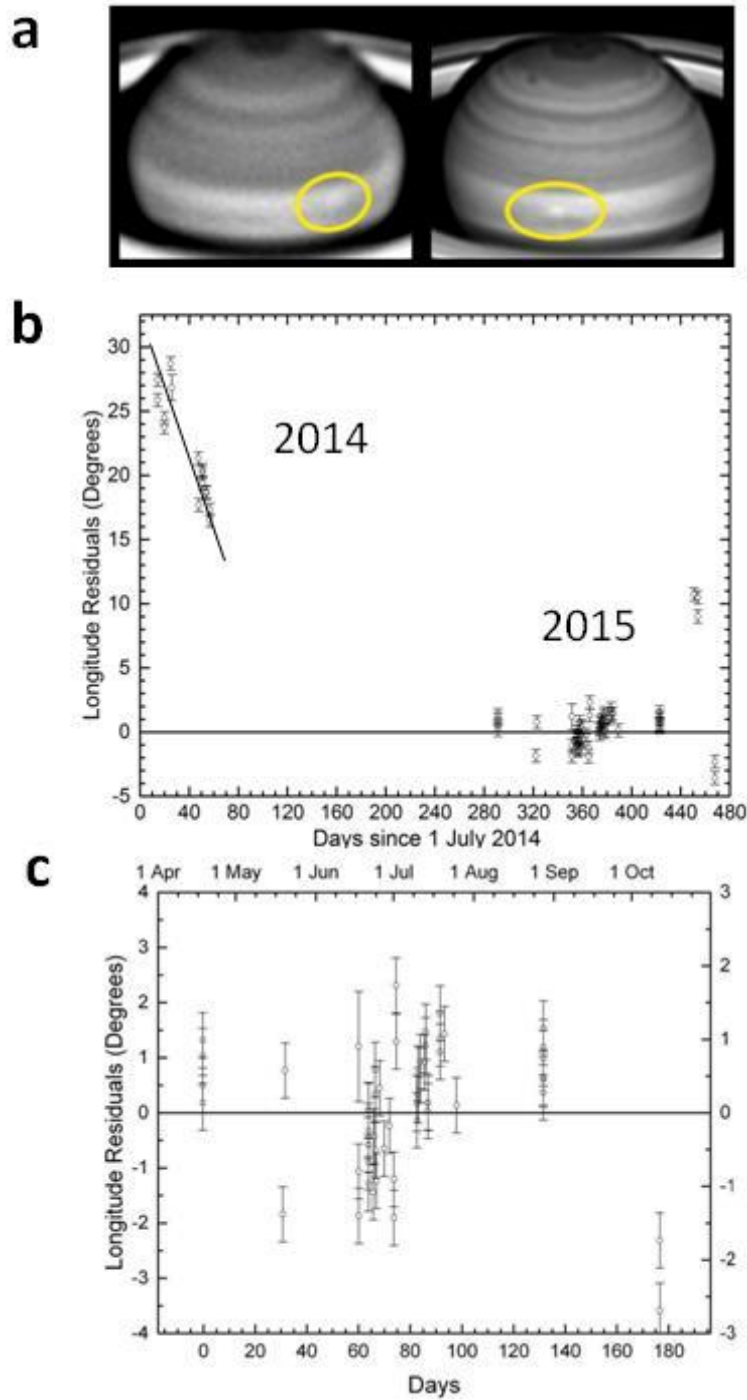

**Figure 2. Ground-based images and motion of Saturn's WS in 2014-2015.** *(a) June 21 (left, Aula EspaZio 28 cm telescope, wavelength 615-950 nm [29]) and July 13 (right, PlanetCam at the 2.2 m telescope Calar Alto Observatory, 750-950 nm [30]); (b) Residuals in the longitude position of spots measured relative to a reference system rotating with a period of 10 hr 11 min 26.3 s (horizontal black continuous line), found from the mean zonal drift of WS between April 18 and October 12, 2015 (-37.05°/day in System III [31] or u = 447 ms$^{-1}$). (c) Detail of the residuals in longitude measured in 2015.*

In order to gain further insight, we observed Saturn with the Hubble Space Telescope (HST) Wide Field Camera 3 (WFC3) on 29, 30 June and 1 July 2015 over a broad spectral range spanning from 255 nm to 937 nm (Supplementary Table 2). A maximum resolution of ~ 260 km/pixel at the sub-observer point is reached in these images. Figure 3 shows maps of the Equatorial Zone at wavelengths sensitive to different altitude levels where the cloud morphology becomes distinct and conspicuous. Images in the continuum band centered at 750 nm shows a variety of features with the WS morphology dominating the scene. The DS spot locates South-East of the WS (equator ward of it). A similar pattern of cloud morphologies, although with a smaller contrast brightness, is visible at 689 nm and 937 nm. In the 890 nm methane absorption band the visible cloud patterns, located higher in the atmosphere, are different to those seen in the red continuum. Dark and white filamentary areas spread along the Equator with a dark region situated close to the DS. In the ultraviolet (336 nm), the contrast between the features is small and reverses in brightness when compared to the 727 nm and 890 nm bands. For example, the dark area in 890 nm becomes brighter than its surroundings at 336 nm. This is consistent with the DS being a region depleted in particles relative to its surroundings, which becomes dark in the methane bands due to gas absorption and bright in the ultraviolet due to the increasing effect of Rayleigh scattering by the gas at shorter wavelengths.

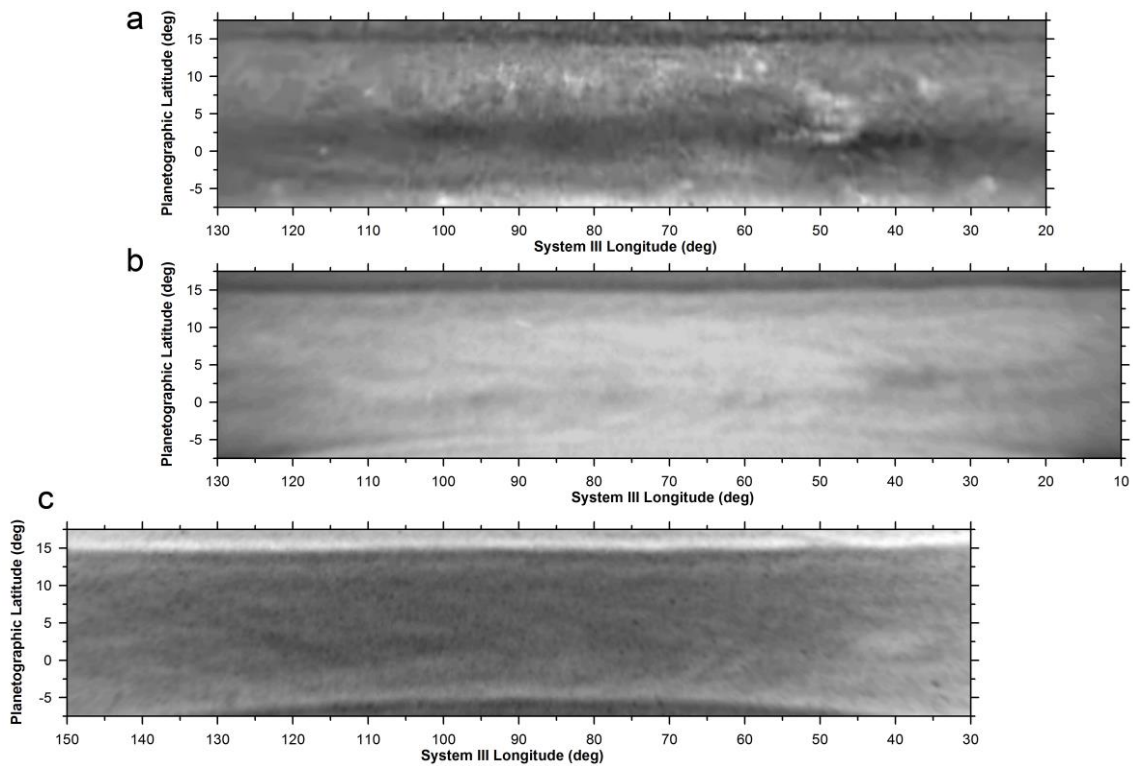

**Figure 3. Maps of Saturn's Equator from Hubble Space Telescope.** *The images were obtained with the WFC3 on June 29, 2015 at the following wavelengths: (a) 750 nm (red continuum); (b) 889 nm (methane band); (c) 336 nm (ultraviolet). Image details are given in Supplementary Table 2.*

Observed at 750 nm at the HST resolution the WS shows its cloud morphology with detail (Figure 4). The WS is a complex feature consisting of a cluster of bright clouds with a size of ~ 300-500 km, extending from latitudes ~ 2°N to 8°N forming the single spot WS observed at ground-based resolution, being its full size ~ 7,000 km. At the HST resolution morphology changes and local motions in the WS are noted in just two Saturn rotations (20.6 hrs). However, the WS preserved globally its coherence during the observing period, in agreement with the long-term ground-based observations. With so few observations and data we cannot assign the dynamical nature of this structure (e. g. if it is a convective storm, some kind of Equatorial wave, or the result of a zonal flow instability).

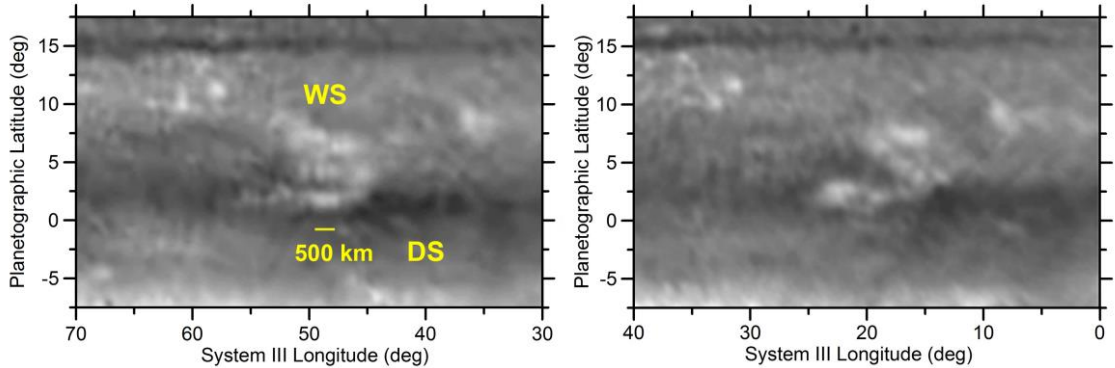

**Figure 4. Morphology of spots WS and DS.** *Morphology and changes in both spots as observed with the HST-WFC3 at a wavelength of 750 nm in images separated by 20.6 hours on June 29$^{th}$ (left) and 30$^{th}$ (right). Image details are given in Supplementary Table 2.*

The Equatorial Zone (EZ) is broad and uniform in reflectivity from latitude ~ 8°S (where the rings projection limits the visibility) to latitude ~ 16°N (where a narrow dark belt locates), in the UV (225 nm), blue (410 nm) and green (502 and 547 nm), as observed with the HST-WFC3 (*Supplementary Table 2*). The WS cloud morphology is similar at the nearby red continuum wavelengths of 689 nm and 937 nm although the contrast between the features decreases at 689 nm when compared to 750 nm (Figures 4 and 5). We will show later that this dark belt is placed at the latitude where the wind profile at the upper haze has a strong change in the velocity. The bland aspect of the EZ at these wavelengths, with almost no contrast between the few detected features, results from the sunlight multiply-scattered in an optically dense haze at depth, quantitatively shown in next section.

**Cloud altitudes.** The ample HST wavelength coverage allows to sense different altitudes within the hazes and clouds. In our radiative transfer model the gas optical depth ($\tau$) is due to Rayleigh scattering by a mixture of $H_2$ plus He and to absorption by $CH_4$ with absorption coefficient $k_{eff}$ [33-35]. The pressure level at which the gas reaches optical depth unity is given in Table 1 and serves as a first step to constrain the altitude of the cloud features as a function of wavelengthor filter employed).

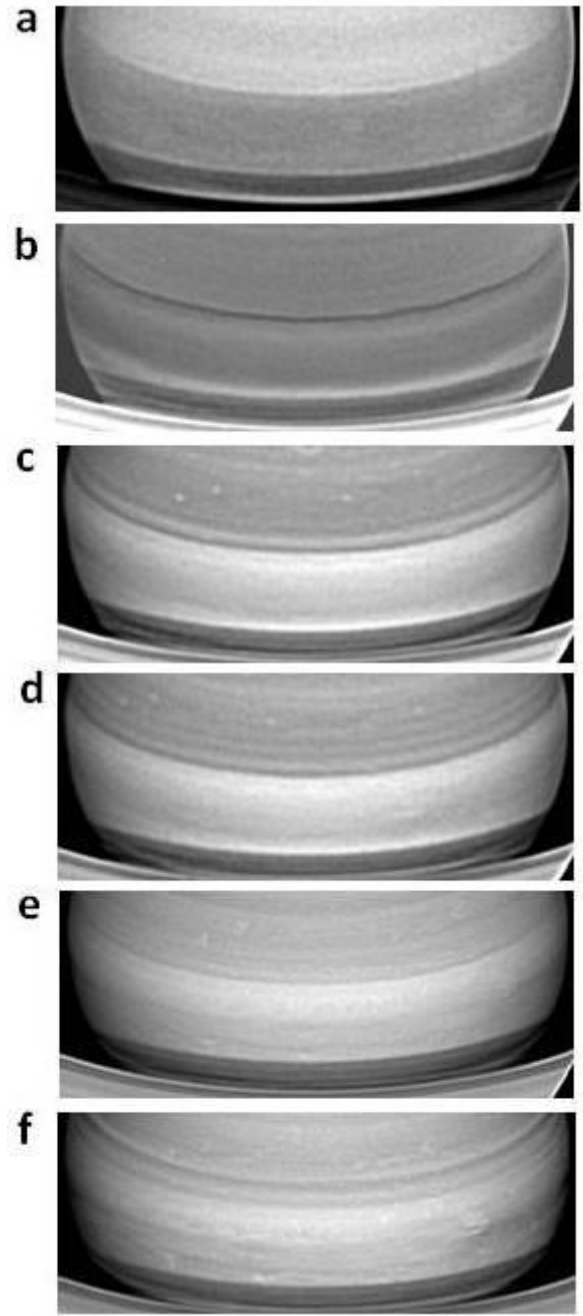

**Figure 5. Multi-wavelength aspect of Saturn's Equatorial Zone.** *HST-WFC3 images obtained on June 29$^{th}$ at the following wavelengths: (a) 225 nm; (b) 420 nm; (c) 502 nm; (d) 547 nm; (e) 689 nm; (f) 937 nm. Image details are in Supplementary Table 2.*

**Table 1: HST Filter list and atmospheric altitude sensitivity**

| Filter | $\lambda_{eff}$ (nm) | $k_{eff}$ (1/km-am) | P ($\tau$=1) (bar) |
|---|---|---|---|
| F225W | 237.8 | $< 10^{-6}$ | 0.112 |
| F336W | 335.9 | $< 10^{-6}$ | 0.515 |
| F410M | 410.8 | $7.2 \times 10^{-5}$ | 1.210 |
| F502N | 501.0 | 0.0017 | 2.760 |
| F547M | 545.2 | 0.0221 | 3.815 |
| F689M | 688.0 | 0.2456 | 5.772 |
| FQ727N | 727.7 | 3.2302 | 0.993 |
| FQ750N | 750.2 | 0.0211 | 13.183 |
| FQ889N | 889.4 | 23.733 | 0.138 |
| FQ937N | 937.7 | 0.0397 | 24.866 |

**Table 2. Cloud altitudes from Radiative Transfer Modeling**

| Layer | Parameter | EZ | WS | DS |
|---|---|---|---|---|
| *Stratospheric haze* | $P_1$ (mbar) | 10±10 | | |
| | $P_2$ (mbar) | 24±12 | | |
| | $\tau_1$ (937nm) | 0.1±0.1 | | |
| | $m_r$ | 1.43 | | |
| | $m_i$ | 0.001 | | |
| | a (µm) | 0.4±0.2 | | |
| | b | 0.1 | | |
| *Tropospheric haze* | $P_3$ (mbar) | 30±10 | | |
| | $P_4$ (mbar) | 300±100 | | 600±300 |
| | $\tau_2$ | 9±4 | | 12±5 |
| | $\varpi_0$ (225nm) | 0.84±0.03 | | |
| | $\varpi_0$ (336nm) | 0.7±0.1 | | |
| | $\varpi_0$ (410nm) | 0.85±0.02 | | |
| | $\varpi_0$ (502nm) | 0.98±0.06 | | |
| | $\varpi_0$ (547nm) | 0.99±0.05 | | |
| | $\varpi_0$ (>689nm) | 1.000 | | |
| | f | 0.8 | | |
| | $g_1$ | 0.8 | | |
| | $g_2$ | 0.3 | | |
| *Bottom cloud* | $P_5$ (bar) | 1.4 | 1.0 ± 0.3 | 5.0±4.0 |
| | $P_6$ (bar) | 1.5 | 2.0 ± 1.0 | 5.0±4.0 |
| | $\tau_3$ | > 10 | | |
| | $\varpi_0$ | 0.998±0.001 | (1.000) | |

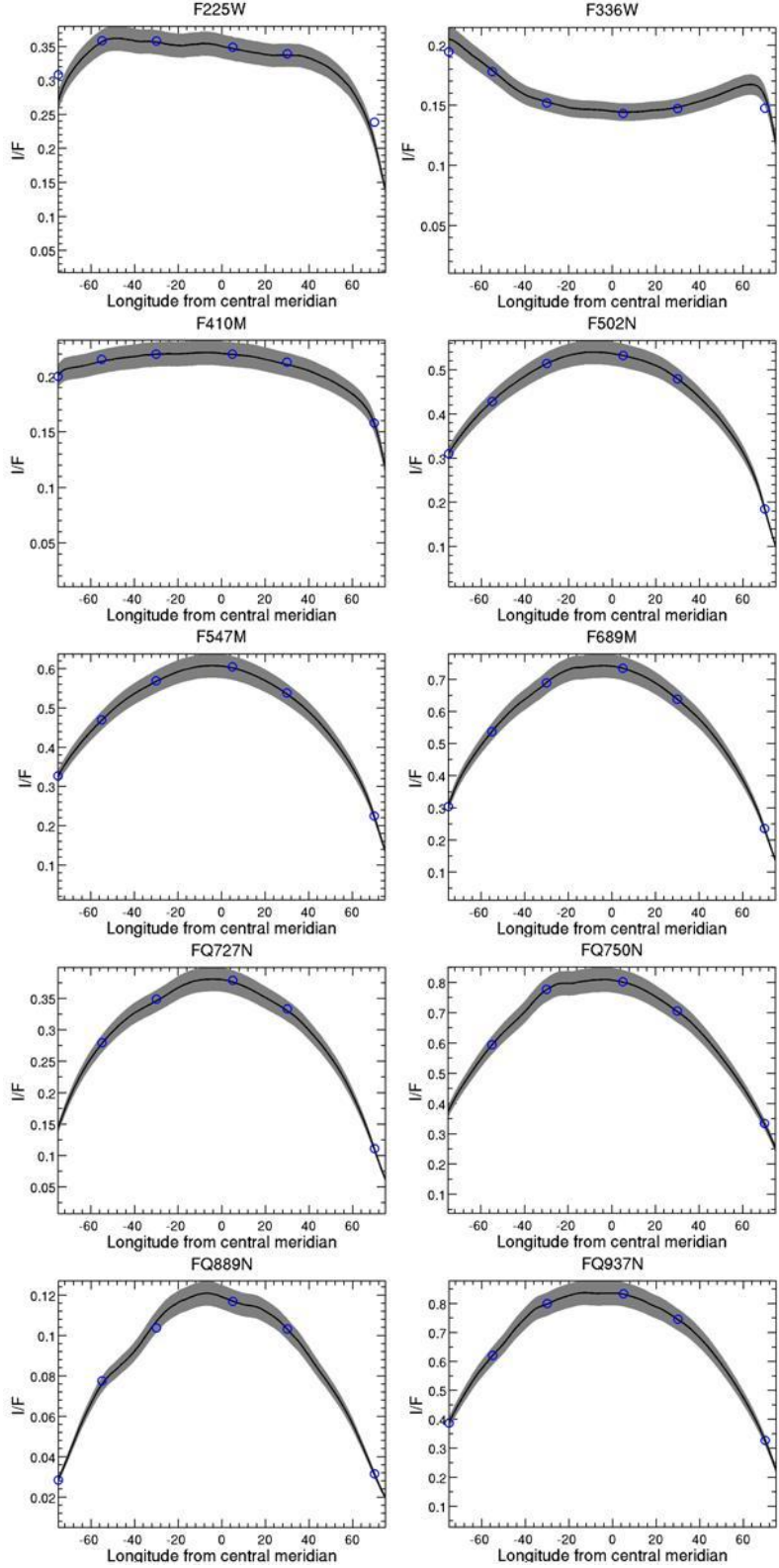

**Figure 6. Radiative transfer models compared to the observed reflectivity.** *Best-fitting model (blue dots) of the EZ center to limb variation of the absolute reflectivity I/F for all HST filters listed in Table 1. The solid lines (with shadowed uncertainties) represent the observed reflectivity and were obtained after smoothing every 1º in longitude to remove particular features from the modeling. Error bars show the 5% confidence level resulting from a number of uncertainty sources* [33-34].

A radiative transfer model was employed to quantify the effect of the particles on the cloud altitudes that we sense at these wavelengths. We model the absolute reflectivity *I/F* (*I* radiant intensity; $\pi F$ the solar flux) determined along latitude circles from limb to terminator (center to limb reflectivity variation, CTLV) in the Equatorial Zone (EZ) at each observed wavelength (see Methods). We used a forward atmospheric model based in a doubling-adding scheme that assumes a plane-parallel atmosphere with three standard separated aerosol layers (from top to bottom, a stratospheric haze, a tropospheric haze and a cloud deck) [33, 35-38]. In Figure 6 we show the best case of the model fit to the observed reflectivity for the EZ and for the bright spot (WS) and dark spot (DS), and in Table 2 we present the resulting model parameters. The best fitting models for the WS and DS features when compared with their measured spectral reflectivity are given in *Supplementary Figure 1*. A sensitivity analysis has been performed for the WS to test the effect on *I/F* when varying the altitude of the bottom cloud as shown in *Supplementary Figure 2*. We have also studied how sensitive the 889 nm reflectivity measurements are to the top pressure and optical thickness of the tropospheric haze around the fitted value given in Table 2 (*Supplementary Figure 3*). All this allowed us to fix the best fit model and the error bars for the parameters given in Table 2.

According to our radiative transfer modeling, the features seen in the wavelength range 336-727-889 nm are located close to the top of a dense tropospheric haze at 60±30 mbar altitude level. Cloud elements observed in the red continuum at 689-750-937 nm, outside the WS feature, are located within the tropospheric haze at altitude levels between 400 - 700 mbar in agreement with our previous works [14, 33]. The WS high brightness requires single scattering albedo $\omega_0 \geq 0.997$ and the cloud top of the feature to be at a pressure $P_{WS} = 1.4 \pm 0.7$ bar.

**Wind measurements.** Tracking the motions of small features in images at different wavelengths on the three dates when the HST images were obtained, allowed us to retrieve the zonal wind profiles (velocity as a function of latitude) at three different altitudes in the peak of the Equatorial jet (latitudes from 20°N to 20°S, Figure 1). Additionally, we measured wind speeds in Cassini ISS images taken in April, September and October 2014 at similar wavelengths, namely red continuum (752 and 939 nm) and the strong methane absorption bands (728 and 890 nm) [39] (see Methods). We then compared those winds with our previously published Cassini ISS wind profiles [12] (Figure 7). First, in the 336-727-890 nm wavelength group the retrieved winds from HST data globally agree with the Cassini profile at 728-890 nm for 2014. The narrow central equatorial jet persists but a meridional broadening seems to have occurred between 2004 and 2009 [12] and during the current epoch 2014-2015 (Figure 7a). Second, at red continuum wavelengths (689-750-937 nm) the HST data shows two groups of velocities (Figure 7b). One group is related to WS and the wind speeds cluster at ~450 ms$^{-1}$ at latitudes from 2N° to 8°N. The individual spots within the complex structure of WS moved with speeds in the range $u$ = 425-475 ms$^{-1}$, agreeing with ground-based data. We assume that all the points that follow these fast motions (black line in Fig. 7b) are located at the same altitude level than that of WS, i.e. deeper relative to the other tracers. Globally, these rapidly moving features captured in HST images match the Voyagers mean profile, also added for comparison in Figure 7b (green line) [7]. Close to latitude 8°S, in a region where wind data from the Voyagers are not available, we find a group of tracers with high velocities. Tentatively, one can identify these features as delineating a southern jet peak, symmetric in latitude to the northern one. If so, Saturn's Equatorial jet at the lowest levels sampled has a similar shape to the Jupiter equatorial jet, i.e. exhibiting a "double symmetric peak" relative to Equator [1, 3-6]. Third, the red continuum 689-750-937 nm wavelengths show a second group of lower velocities that follow the 2014 profile from Cassini ISS (blue and violet lines, Fig. 7b). This profile (Cassini 2014 and HST 2015) exhibits a velocity shift by ~ +20 ms$^{-1}$ above those measured with Cassini ISS in 2004-2008 (red line, Fig. 7). The presence in 2015 of the two different velocity profiles at red continuum wavelengths can be interpreted as the detection of clouds moving at two different altitude levels. According to radiative transfer modeling the profile showing low velocities corresponds to the 400-700 mbar level whereas the profile with higher velocities corresponds to the 1.4 bar level (in agreement with the WS velocity). We rule out waves as an explanation for the dual profile since no periodic structures are seen on the albedo patterns (Figure 3). Finally, as a further comparison, we have included in Figure 7b a smoothed version of the wind profile derived from 5-μm images obtained by Cassini VIMS [32] that senses cloud features as opacity sources to the infrared emission from the planet at deeper levels (1-3 bar) in the atmosphere [40-41].

**Vertical wind shear of the zonal winds.** The above results can be used to retrieve the vertical structure of the zonal wind in the peak of the Equatorial jet in 2014-2015 at cloud level. We selected representative latitudes between 15°N and 10°S to draw the zonal velocity $u$ as a function of altitude as shown in Figure 8a where for completeness we also include published data on the deeper winds from Cassini VIMS [32]. Outside the latitude band from ~ 15°N to 15°S the zonal winds exhibit low vertical shears (see also Fig. 1 and 7). The largest vertical wind shears are concentrated in the latitude band from 10°S to 10°N and in altitude between the tropospheric haze and the cloud deck (from levels 0.5 bar to 1 - 4 bar). The most intense shears occur in the equatorially symmetric jets at latitudes ~ 5°S and 5°N where $\partial u/\partial z = 100$ ms$^{-1}$/H = 2.6 \times 10^{-3}$ s$^{-1}$ (for $H = 40$ km being the scale height) or ~ 2.5 ms$^{-1}$/km (2.5×10$^{-3}$ s$^{-1}$). This shear is two orders of magnitude higher than the maximum meridional shear of the zonal flow found at Saturn's Equator $\partial u/\partial y = 3 \times 10^{-5}$ s$^{-1}$ (Figures 1 and 7).

In order to explore the long-term behavior of the vertical profile we have combined all the available wind data gathered during the last 35-year period (1980-2015) i.e. over ~ one Saturn year (29.5 years). We concentrate into a single latitude band from 4°N to 8°N, i.e. the latitudinal range occupied by WS which is also the latitude where the 1990 GWS developed [19-21, 42] (Figure 8a). A conclusion we reach is that large changes in the speed of the zonal flow occurred above the ~ 1 bar altitude level. We have also taken into account retrieved thermal winds available for the Voyager and Cassini periods at these latitudes and altitudes [43] (Figure 8b). When considering the whole data set we first conclude that winds dropped between the 1980-81 and 1990-91 (following the GWS storm eruption) by ~ 150 ms$^{-1}$ at the cloud top level (~ 60 mbar) and by ~ 100 ms$^{-1}$ at mid-altitude levels (~ 350 -700 mbar). Second, at the top level a wind speed increase of ~ 50 ms$^{-1}$ is detected between 1996-2008 and 2014-15 (Figure 8b). Third, at the mean altitude level winds kept nearly constant (Figure 8b) after 1990-91, but an intensification of the jet speed by ~ 25 ms$^{-1}$, preserving its meridional structure, occurred between 2004-2008 and 2014 at ~ 500 mbar (latitudes 10°S and 10°N) (Figure 8b). We cannot determine if this is a real change or simply a vertical shift in the altitudes where the tracers are located within the tropospheric haze (this will correspond to a downward shift of the tracers by about one scale height). Finally, temporal variability at deeper levels (below ~ 1.4 bar) cannot be constrained with the present data. New wind measurements using VIMS images at 4-5 microns are necessary to sound this level.

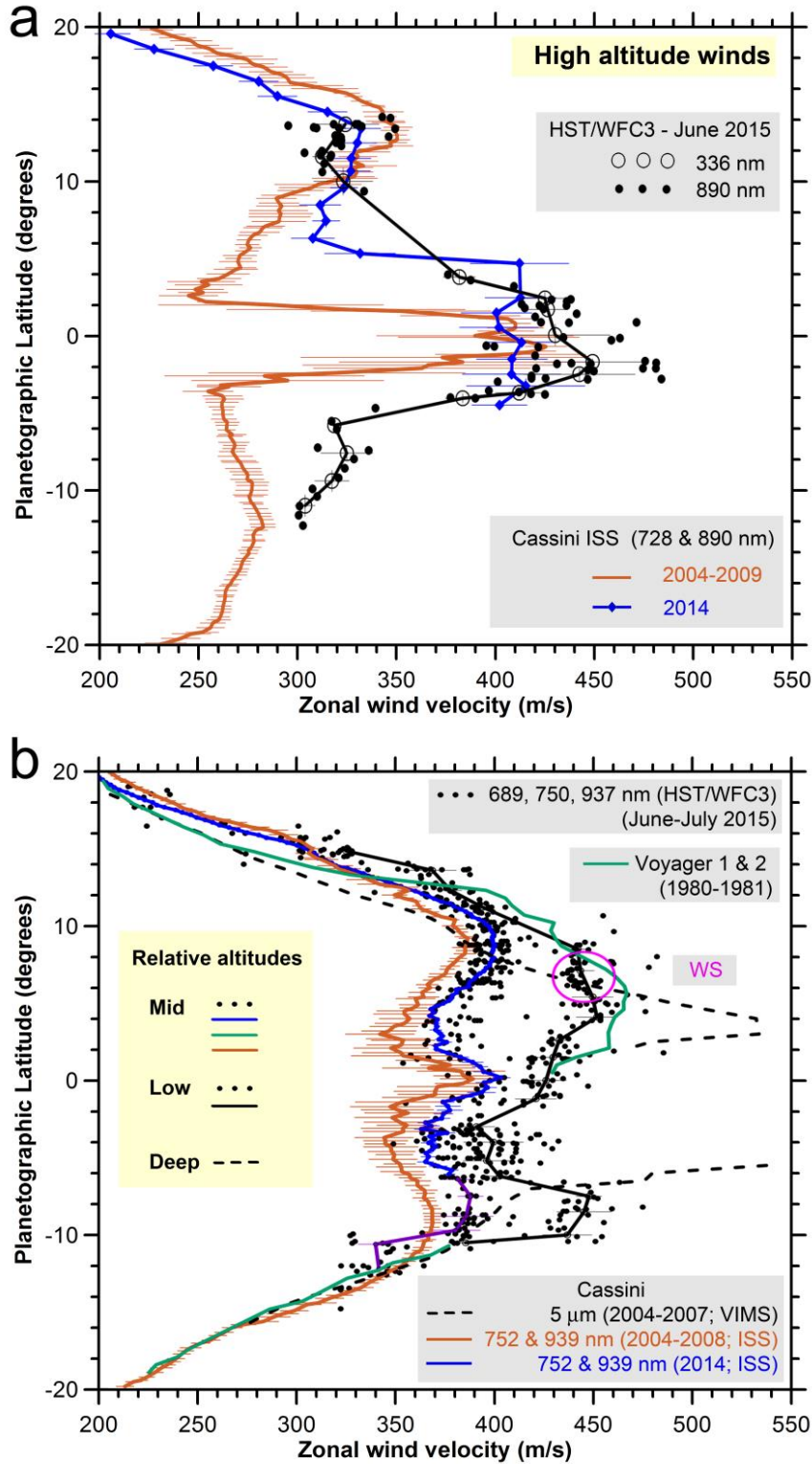

**Figure 7. Saturn's Equatorial zonal wind profiles at three altitude levels.** *(a) High altitude winds for April to October 2014 (blue line) and June 2015 (black line), compared to an average from 2004 to 2009 (magenta line, [12]). (b) Mid level winds (violet line and blue profile) and low altitude winds (black line) in 2014-2015. The violet and black lines are fits to the two groups of data points. They are compared to previous wind data from 1980-1981 [7] (turquoise line), 2004 to 2007 [32] (deeper winds, dashed line) and 2004-2009 [12] (magenta line). The gaps are due to rings projection and shadow. The 2015 WS features are those within the circle.*

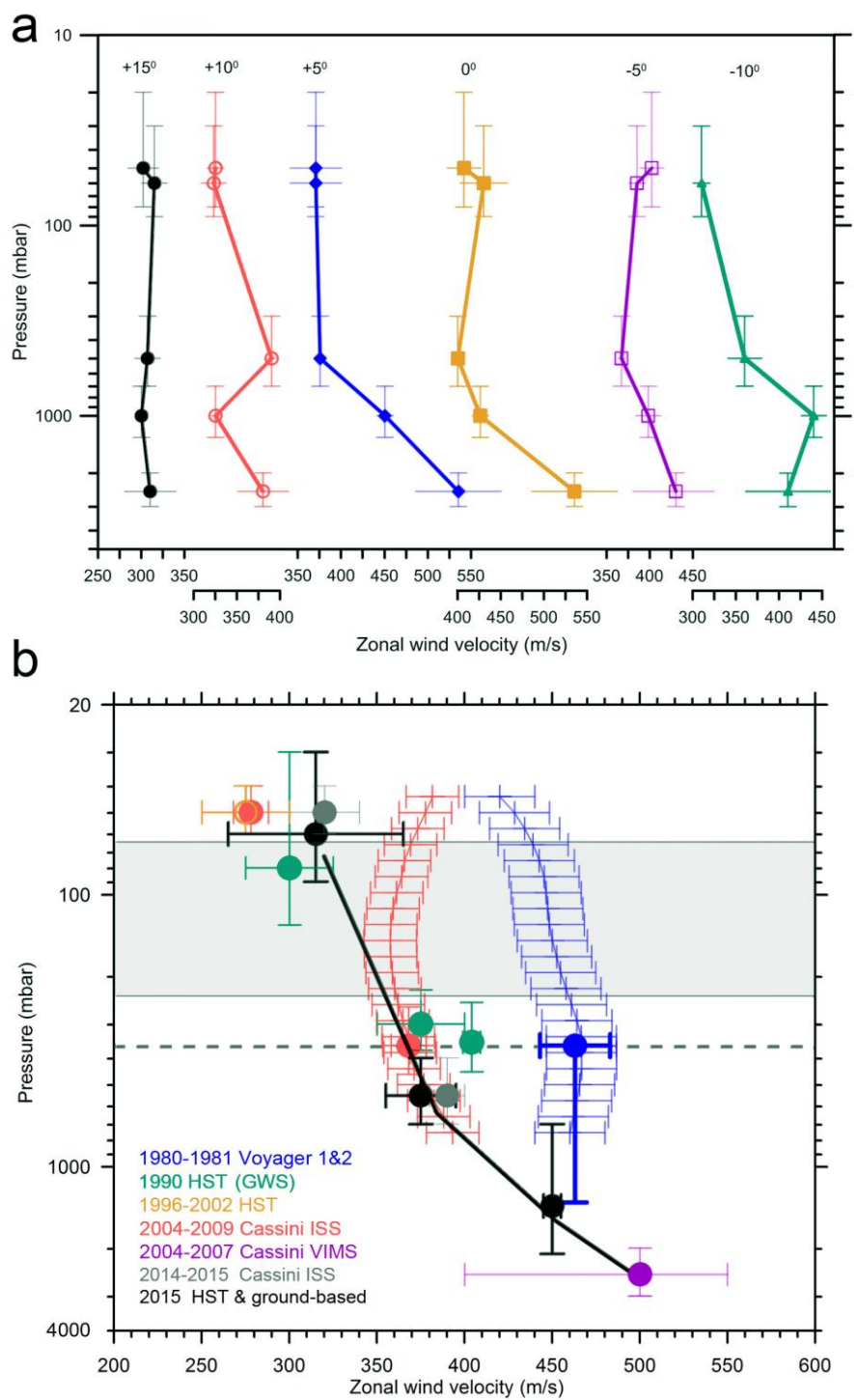

**Figure 8. Vertical structure and temporal variability of Saturn's Equatorial winds.** *(a) Vertical profiles of the zonal wind retrieved in 2014-2015 at 6 planetographic latitudes. Altitude retrievals correspond to the following wavelengths: 50 mbar (890 nm, dots in Fig. 7a), 60 mbar (336 nm, circles in Fig. 7a), ~500 mbar (689-750-937 nm, blue-violet line in Fig. 7b), ~1.4 bar (689-750-937 nm, black line in Fig. 7b), ~2.5 bar (VIMS data from [18]). (b) Vertical profiles of the zonal wind (planetographic latitudes +4° to +8°) between 1980 and 2015. Dots are for cloud tracers in 2014-15 (gray and black, dark line) and for 1980-2009 in color [8, 7-9, 10, 12, 19-20, 32, 39]. Color lines are from thermal wind retrievals [13, 43] with wind reference level shown by a dashed horizontal line (360 mbar). The shaded area indicates the location of the equatorial tropopause [44].*

**Discussion**

Our results indicate that Saturn's broad Equatorial jet, spanning from ~35ºN to 35ºS (Figure 1), has a central peak (from latitudes ~ 10º N to 10ºS) that has an intense vertical wind shear in its upper cloud and haze layers (from the tropopause at ~ 0.06 bar to the upper cloud at ~ 1-2 bar) and that has experienced temporal variability in the velocity of the upper cloud and haze levels. There are at least three known sources that can be involved in the variability and vertical structure. First, three of the six rare Saturn giant planetary-scale GWS storms have originated in the peak of the equatorial jet in 1876, 1933 and 1990 [2, 19-21], the last one accompanied by large activity in 1994 [42, 45]. They probably represent one of the sources for major and long-lasting changes in the area although the operating mechanisms are not yet well understood [2, 46]. New "shallow water layer" dynamical simulations of the 1990 GWS show the effect of changing the equatorial jet structure on the storm evolution (therefore fixing the jet structure at the epoch of the storm's outbreak) and the generation of abundant Rossby waves affecting the structure of the jet peak [47]. The upward propagation of this wave activity could have formed "an equatorial beacon" in the upper stratosphere similar to that observed in the GWS 2010 [48-49].

Second, since the solar radiation penetrates down to altitudes ~ 2 bars [18], variability in the vertical upper cloud and haze distributions [33-38] coupled to seasonal radiative effects enhanced at these latitudes by ring-shadowing is the most plausible second source for changes [50]. Seasonal variability is observed in the temperature field above ~ 400 mbar [40-41]. The 2004-2014 Cassini temperature retrievals shows that aerosol heating produces a "kink" in the vertical temperature profiles from 100 to 400 mbar in the Equator [40]. This indicates that variability should occur in the thermal winds as retrieved from a modified thermal wind balance equation at Equator [51], although large uncertainty exists in these retrievals [40, 43, 49] (Figure 8b). However, radiative-dynamical modeling points to the seasonal variability and vertical wind shear occurring in the peak of the jet just at the right latitudes and altitudes we observed [50, 52]. A three-dimensional Outer-Planet General Circulation Model (OPGCM) applied to Saturn showed that a meridional circulation develops at Equator and was found to be dominated by a seasonally reversing Hadley circulation [52]. The temporal changes are predicted to occur at altitudes above ~ 200 mbar, with the reversion of the Hadley circulation taking place at latitude ~ 25°. In the altitude range ~ 55 – 200 mbar the model generates a symmetric jet at both sides of the Equator (latitudes ~ 8°N and 8°S) with a seasonally variable speed of ~ 60 ms$^{-1}$ and whose maximum value is ~ 350 ms$^{-1}$. Maximum vertical shears in this altitude range predicted by the model are ~ +0.6 ms$^{-1}$ km$^{-1}$ (6.2x10$^{-4}$ s$^{-1}$) agreeing with other model predictions [50]. These models doesn't exactly match the observations (for example they do not reproduce the central jet observed at 60 mbar) but they show, in agreement with the observations, that the velocity changes forced by the seasonal insolation cycle occur in the center of the Equatorial jet at the expected altitudes (~ 55 – 200 mbar).

The Semi-Annual Oscillation (SAO) that has been detected in the stratosphere at altitudes above 20 mbar [25-26] could be the third mechanism involved in the velocity changes affecting winds at least down to the top of the upper haze at ~ 60 mbar. We suspect that the changes at 60 mbar observed in 2004-2009 and 2014-2015 might be related to the SAO cycle. However, the observed downward propagation of the SAO damps at this level [44, 53]. The lack of measurements of the amplitude and vertical extent of the SAO below 20 mbar does not allow searching for its implication or coupling with the wind changes we observe at 60 mbar.

Our results show that in order to constrain the coupling of the SAO (as derived from temperature measurements) to the wind field (as measured by cloud tracking) in the upper troposphere (50 – 200 mbar) we need more long-term simultaneous observations of both magnitudes, along with the characterization of the vertical structure of cloud and hazes where the winds are measured. Additionally, to characterize the vertical structure, it would be highly desirable to accompany these data with long-term wind measurements at deeper levels (1-3 bar) using the 5 micron window as in Cassini VIMS images [32]. Theoretical modeling will probably require a refinement of the GCM models [52] that should be able to capture the SAO cycle and vertical structure of the Equatorial jet across the stratosphere and upper troposphere. For example they should be able to reproduce the intense jet centered at the Equator as observed at 60 mbar and the mechanisms that transport and concentrate momentum in such narrow latitude band. Latent heat release by water condensation at the level $P \sim 10$ bar in Saturn [54], and radiative relaxation effects on the zonal flow have been proposed as elements in the generation of such narrow Equatorial jets in shallow layers [55]. Another important issue in modeling the upper troposphere dynamics is the imposed lower wind boundary condition, i. e. how winds behave below the accessible region to remote observations ($P \geq 3$-4 bar). For example models of deep extending winds under deep forcing [56-58] reproduce the double jet peak we observe in the Equator of Jupiter [1] and Saturn (figure 7, [32]). Radiative-dynamical models [51-52] should take this into consideration.

To complete the research it would be very interesting to explore with the GCM [52] how the dynamics of the Equatorial great storms (the GWS phenomenon) influenced both the stratospheric SAO and the global wind field in the region. This could be done injecting a heat or mass source at the appropriate latitude [47] A reanalysis of the available infrared data for the GWS 1990 [25] and the following up activity across 1994-1995 will serve to search for "beacon" activity in equatorial storms.

**METHODS**

**Image navigation and wind velocity measurements.**
Ground-based observations and HST and Cassini ISS images were navigated for limb position and features location with software LAIA and PLIA [59-60]. We used three different techniques to retrieve wind speeds: (a) Cloud-tracking of individual features; (b) Latitudinal correlation of longitudinal brightness patterns; (c) Supervised two dimensional brightness correlation [61-62]. We used image pairs separated by 20.6 hr but in some cases for method (a), tracking was possible in three images with time separations of ~ 20 hrs between each pair. Typically we resolve cloud elements with a size of ~ 500 km giving a formal error of ~ 7 ms$^{-1}$. The data points were binned in boxes with a width of 1°-2° in latitude to get zonal profiles and then the mean values, with standard deviation of ~ 10 ms$^{-1}$, joined to simple straight lines.

**Photometric calibration.**
HST observations were photometrically calibrated following the WFC3 handbook instructions [63]. Radiances were converted into absolute reflectivity I/F using the solar spectrum [64-65]. The resulting I/F as a function of planetary geographical coordinates was confirmed against values given by other authors for selected locations of the planet and with geometric albedo values of the planet [34].

**Radiative transfer model.**
We selected center to limb scans along latitude 5°N where the bright spot resides. However, we extrapolated these results to the observable Equatorial Zone (latitude range between 15°N and 10°S) in view of the similar photometric behavior at all wavelengths. The forward modeling allows retrieving the haze and aerosol vertical structure in the region down to the ammonia ice cloud at 1.4 bar. The reference vertical structure we found is similar to those retrieved in previous works [11-33]. The forward model has been described in previous papers [33, 36]. It assumes a plane-parallel atmosphere and it is based in a doubling-adding scheme. The model includes gaseous absorption by $CH_4$ and scattering by a mixture of $H_2$ and He. Methane absorption coefficients were calculated through a convolution of the system throughput curves with methane absorption spectrum [35]. The model also accommodates particle absorption and scattering by means of a number of possible phase functions. For the altitude location of the two particular fast moving features (WS and DS) we consider two model scenarios: higher single scattering albedo $\omega_0(\lambda)$ and different cloud top pressure level ($P$) of the feature, always within the tropospheric haze. Under this model, the WS high brightness gives $\omega_0 = 1$ and puts the cloud top of the feature down to $P_{WS} = 1.4 \pm 0.7$ bar. For the dark feature, instead, the model indicates a lower particle density due to a deeper location of the tropospheric haze bottom.

**Inversion of atmospheric parameters.**
In order to determine the atmospheric parameters that most likely reproduce the observed reflectivity as a function of wavelength and scattering angles we computed the mean square deviation for each model. The goal was to minimize this function in the multi-dimensional space of free parameters, for doing so we used a Nelder-Mead simplex method [66]. A number of runs were performed from some initial states in the free parameter space to ensure that the retrieved minimum of the function was the absolute minimum in the range of confidence. Sensitivity and error bars are explored uni-parametrically following previous works [67]. An example of the exploration of model sensitivity for some key parameters is shown in SI.

**Code availability.**
The codes for radiative transfer models are available on request from S.P.H. This includes the forward model (FORTRAN 77), high-level interfaces for plotting and interacting with the minimization routines and a trivial call to native simplex algorithms (Python and MATLAB).

Details about the HST images and observations from the PVOL contributors are listed in the Supplementary Information.

**Acknowledgements**
This work is based on observations and analysis from Hubble Space Telescope (GO/DD program 14064), Cassini ISS images (NASA pds), and Calar Alto Observatory (CAHA-MPIA). A.S.-L. and UPV/EHU team are supported by the Spanish projects AYA2012-36666 and AYA2015-65041-P with FEDER support, Grupos Gobierno Vasco IT-765-13, Universidad del País Vasco UPV/EHU program UFI11/55, and Diputación Foral Bizkaia (BFA).

**Author Contributions**
A.S.-L. performed the wind analysis and wrote the manuscript; A.S.-L., E.G.-M., R.H., A.S., A.A., N.B.-I., I.G.-L., J.M.G.-F., performed the wind measurements on Cassini and HST images; A.S.-L., M.H.W., A.S., S.P.-H., J.F.-R., T.dR.-G., R.H., I.dP., prepared the HST observations. S.P.-H., J.F.S-R. performed the radiative transfer analysis. L.L. calculated thermal winds. T. B. contributed a large number of the PVOL ground-based images. All the authors discussed the results and commented on the manuscript.

**Competing financial interests**
The authors declare no competing financial interest

**Additional Information**
Supplementary is available in the online version of the paper.
Correspondence and requests for materials should be addressed to A.S-L.